\documentclass[12pt]{iopart}
\usepackage{indentfirst}
\usepackage{cite}
\usepackage{booktabs}
\expandafter\let\csname equation*\endcsname\relax
\expandafter\let\csname endequation*\endcsname\relax
\usepackage{amsmath,amssymb}
\usepackage{graphicx}
\usepackage{subfigure,lscape}
\usepackage{multirow}

\begin{document}
\title{Improving Pre-movement Pattern Detection with Filter Bank Selection}
\author{Hao Jia$^1$, Zhe Sun$^{2*}$, Feng Duan$^{3*}$, Yu Zhang$^{4*}$, Cesar F. Caiafa$^{5*}$, Jordi Sol{\'e}-Casals$^{1*}$}

\address{$^1$Data and Signal Processing Research Group, University of Vic-Central University of Catalonia, Vic, Catalonia}

\address{$^2$Computational Engineering Applications Unit, Head Office for Information Systems and Cybersecurity, RIKEN, Saitama, Japan}
\address{$^3$College of Artificial Intelligence, Nankai University, Tianjin,China}
\address{$^4$Department of Bioengineering, Lehigh University, Bethlehem, PA 18015, USA}
\address{$^5$Instituto Argentino de Radioastronom\'{i}a, CONICET CCT La Plata/CIC-PBA/UNLP, V. Elisa, Argentina}
\ead{zhe.sun.vk@riken.jp, duanf@nankai.edu.cn, yuzhang@lehigh.edu, ccaiafa@gmail.com, jordi.sole@uvic.cat}

\maketitle
\begin{abstract}
Pre-movement decoding plays an important role in movement detection and is able to detect movement onset with low-frequency electroencephalogram (EEG) signals before the limb moves. In related studies, pre-movement decoding with standard task-related component analysis (STRCA) has been demonstrated to be efficient for classification between movement state and resting state. However, the accuracies of STRCA differ among subbands in the frequency domain. Due to individual differences, the best subband differs among subjects and is difficult to be determined. This study aims to improve the performance of the STRCA method by a feature selection on multiple subbands and avoid the selection of best subbands.

This study first compares three frequency range settings ($M_1$: subbands with equally spaced bandwidths; $M_2$: subbands whose high cut-off frequencies are twice the low cut-off frequencies; $M_3$: subbands that start at some specific fixed frequencies and end at the frequencies in an arithmetic sequence.). Then, 
we develop a mutual information based technique to select the features in these subbands. 
A binary support vector machine classifier is used to classify the selected essential features.

The results show that $M_3$ is a better setting than the other two settings. With the filter banks in $M_3$, the classification accuracy of the proposed FBTRCA achieves 0.8700$\pm$0.1022, which means a significantly improved performance compared to STRCA (0.8287$\pm$0.1101) as well as to the cross validation and testing method (0.8431$\pm$0.1078).

\end{abstract}
\vspace{2pc}

\section{Introduction}
\label{sec:intro}
The movements of human limbs arouse potential changes on the human scalp, which can be observed with noninvasive brain-computer interface-based electroencephalogram (EEG) signals \cite{schalk2004bci2000,Ramadan2016Brain}. In studies on the movement detection with EEG signals, motor imagery (MI) is one of the most frequently used brain activities in the motor cortex \cite{padfield2019eeg,zhang_eeginception_2021,zhang2021survey}. When the limbs begin moving, the power of EEG signals in alpha rhythm (frequency range: 8$\sim$12 $Hz$) and beta rhythm (frequency range: 13$\sim$30 $Hz$)  shows an upward or downward trend, which is called event-related desynchronization/synchronization. Because of this phenomena when humans imagine that the left limb or right limb moves, the power changes in the left/right half of the scalp have the opposite reaction. However, power changes in MI occur after limb moves, i.e., movement can only be detected after movement onset in MI analysis. The brain activity, movement-related cortical potential (MRCP), can be used to judge the movement or resting state of human limbs before movement onset. Hence, MRCP is expected to help enhance restoration of useful motor function and reduce the time delay of movement detection \cite{mcfarland_effects_2015,sburlea_continuous_2015}.

MRCP is a kind of low-frequency EEG signal (frequency range: 0.5$\sim$10 $Hz$) acquired in the motor cortex \cite{olsen_paired_2018,mammone_deep_2020,shakeel_review_2015,karimi_detection_2017,wang_enhance_2020}. MRCP analysis is applied to EEG signals located around the movement onset. The readiness potential section is the stage that starts from 2 seconds before movement onset and ends at movement onset. The movement monitoring potential section is the stage that starts from movement onset and ends at 1 second after movement onset. With EEG signals in both the readiness potential section and movement monitoring potential section, Jeong \textit{et al.} proposed the subject-dependent and section-wise spectral filtering (SSSF) method to extract the amplitude features in MRCP and successfully classified the two-class problem between movement and resting states \cite{jeong_decoding_2020}. To optimize the selected features in SSSF, Jeong \textit{et al.} adopted a cross validation and testing (CVT) method to select the best frequency range for each subject. Jia \textit{et al.} proposed a simple but efficient method, standard task-related component analysis (STRCA) \cite{duan_decoding_2021}. The method used spatial filter task-related component analysis (TRCA) and extracted the canonical correlation pattern (CCP) from the filter EEG signals. The STRCA method achieved an improved classification performance with EEG signals only in the readiness potential section. However, STRCA faces the frequency range selection problem when decoding pre-movement patterns in MRCP analysis. Considering the CVT used in SSSF, STRCA can be further improved with the filter bank technique.

The filter bank technique aims to solve the feature selection problem among various subbands in the frequency domain. It is widely used in the analysis of brain activities such as MI and steady-state visual evoked potential (SSVEP). In SSVEP analysis, the canonical correlation analysis (CCA) method is a classical method for detecting stimulus frequencies \cite{bin2009online}. CCA can measure the similarity between EEG signals and the reference signals, and many methods in SSVEP analysis are developed based on CCA \cite{lin2006frequency,Xu2018,jiao2018novel}. Filter bank canonical correlation analysis was proposed to incorporate harmonic and fundamental frequency components, which improved the detection of standard CCA in SSVEP \cite{Chen_2015}. Without the filter bank technique, the CCA faces the problem of selecting frequency components. In MI analysis, the common spatial pattern (CSP) method is the most classical method \cite{pfurtscheller2001motor}. The CSP method consists of the spatial filter CSP and the variance features. The CSP method shows a varying accuracy among the subbands in alpha (8$\sim$12 $Hz$) and beta (13$\sim$30 $Hz$) rhythms \cite{wang2021personalized,padfield2019eeg}. Filter bank CSP (FBCSP) is an advanced MI analysis method that was developed by combining the CSP method and the filter bank technique \cite{Ang2008}. Equipped with the filter bank technique, FBCSP can avoid the selection of subbands and achieves a better and more stable accuracy than the standard CSP method.

In both MI analysis and SSVEP analysis, the filter bank technique used a feature selection method to optimize the extracted CSP features or CCA features in each subband. The best frequency range of the filter bank is different among the subjects due to the individual differences. The feature selection method overcomes the frequency range selection problem and enables the classification to achieve a stable and overwhelming result. When applying the filter bank technique to STRCA, there are three problems to tackle:

\noindent (1) How to select the start and stop frequencies of the subbands in the filter bank technique, i.e., frequency range setting is unknown;

\noindent (2) The feature selection method for STRCA is undetermined;

\noindent (3) The feature arrangement is unclear when applying the feature selection method on STRCA features extracted from all subbands.

This study aims to analyse how to incorporate the filter bank technique in STRCA. To address the above three problems, two steps are adopted in this study. First, three feature range settings are compared to decide how to select frequency range of each subband in pre-movement decoding. Second, a new filter bank TRCA (FBTRCA) method is proposed to decode the pre-movement patterns.

In Section \ref{sec:methd}, the EEG dataset used, the data preprocessing process and the FBTRCA method are introduced. In addition, the proposed FBTRCA method is given in this section. Section \ref{sec:resul} analyses the proposed method from both the frequency range settings and the feature selection methods. Section \ref{sec:conclu} concludes the work in this study and Section \ref{sec:discuss} discusses how the FBTRCA is designed and the shortcomings of FBTRCA.

To facilitate the understanding on the contents of this work, the abbreviations in this work are given in Table \ref{tab:1.1}.

    \begin{table}[htbp]
        \centering
        \tiny
        \label{tab:1.1}
        \caption{Descriptions of Abbreviations}
        \begin{tabular}{c|c|c}
             \toprule
             Abbreviation& Full Name & Description  \\
             \midrule
             EEG & Electroencephalogram & Multi-channel signals acquired from the surface of brain scalp.\\
             MRCP & Movement-Related Component Potential & A kind of brain activity related to pre-movement.\\
             MI & Motor Imagery & A kind of brain activity related to movement.\\
             SSVEP & Steady State Visual-Evoked Potential & A kind of brain activity evoked by visual stimulus.\\
             \midrule
             CCA & Canonical Correlation Analysis & A basic classification method in SSVEP \cite{bin2009online}.\\
             FBCCA & Filter Bank Canonical Correlation Analysis& A method that optimizes CCA by filter bank selection \cite{Chen_2015}.\\
             CSP & Common Spatial Pattern& A basic classification method in MI \cite{pfurtscheller2001motor}.\\
             FBCSP & Filter Bank Common Spatial Pattern& A method that optimizes CSP by filter bank selection \cite{Ang2008}.\\
             SSSF & Subject-dependent and section-wise spectral filtering & A binary classification method for movement and resting states \cite{jeong_decoding_2020}.\\
             \midrule
             STRCA & Standard Task-Related Component Analysis & A binary classification method for movement and resting states \cite{duan_decoding_2021}.\\
             \textbf{FBTRCA} & Filter Bank Tasked-Related Component Analysis & A method that optimizes STRCA by filter bank selection.\\
             
             \midrule
             
             \multirow{2}{*}{\textbf{CVT}} &  \multirow{2}{*}{Cross Validation and Testing} & A method that optimizes STRCA by finding the best \\
              & & frequency range with cross validation and testing.\\
             \midrule
             TRCA & Task-Related Component Analysis & The spatial filter used in STRCA \cite{duan_decoding_2021}.\\
             CCP & Canonical Correlation Pattern & The extracted features in STRCA \cite{duan_decoding_2021}.\\
             \midrule
             MIQ & Mutual Information Quotient & A feature selection method based on mutual information\cite{ding2003miq}.\\
             MAXREL & Maximum Relevance & A feature selection method based on mutual information\cite{peng2005mrmr}.\\
             MINRED & Minimum Redundancy & A feature selection method based on mutual information\cite{peng2005mrmr}.\\
             MRMR & Minimum Redundancy Maximum Relevance & A feature selection method based on mutual information \cite{peng2005mrmr}.\\
             QPFS & Quadratic Programming Feature Selection& A feature selection method based on mutual information \cite{lujan2010qpfs}.\\
             CIFE & Conditional Infomax Feature Extraction & A feature selection method based on mutual information \cite{lin2006cife}.\\
             CMIM & Conditional Mutual Information Minimization& A feature selection method based on mutual information \cite{herman2013cmim}.\\
             MRMTR & Maximum Relevance Minimum Total Redundancy& A feature selection method based on mutual information \cite{nguyen2014mrmtr}.\\
             \midrule
             SVM & Support Vector Machine & A binary classifier \\
             LDA & Linear Discriminant Analysis & A binary classifier \\
             NN & Neural Network & A binary classifier\\
             \bottomrule
        \end{tabular}
        Note: The bold abbreviations indicate that this method is first adopted or proposed in this work.
    \end{table}
    

\section{Material and Method}
\label{sec:methd}
\subsection{Dataset Description}
The dataset used in this study is a public dataset. Considering that the proposed FBTRCA method is developed based on STRCA, the same EEG dataset and preprocessing procedure in STRCA is adopted when processing the EEG signals \cite{duan_decoding_2021,ofner_upper_2017,ofner_attempted_2019}. The dataset consists of 7 states with 15 subjects. The 7 states include the resting state \textit{rest} and 6 movement states \textit{elbow flexion}, \textit{elbow extension}, \textit{supination}, \textit{pronation}, \textit{hand close} and \textit{hand open}. The EEG signals used were acquired from 11 channels with active electrodes. These channels are located around the motor cortex. According to 10/20 international system, 5 electrodes of the 11 electrodes are located at the center of motor cortex: $FC_z$, $C_3$, $C_z$, $C_4$, $CP_z$; and 6 electrodes are located at the surrounding of motor cortex: $F_3$, $F_z$, $F_4$, $P_3$, $P_z$, $P_4$. The ground electrode was placed on $AF_z$ and the reference electrode was placed on the right mastoid. The EEG signals are bandpassed with an 8-order Chebyshev bandpass filter from 0.01 $Hz$ to 200 $Hz$. The sample rate of EEG signals is 512 $Hz$ but the signals are downsampled to  256 $Hz$ considering the computation load. A notch filter at 50 Hz is applied to avoid the influence of power line interference. The movement trajectory $Hand_y$ acquired with a glove sensor, which is used to locate the movement onset of limb movement.

In the acquisition paradigm of the dataset, a trial lasts five seconds. At the start of a trial, the computer screen displays cross and beep sounds. Two seconds later, the computer screen shows a cue that indicates the required movement (or resting). The trial of each motion repeats randomly 60 times for 15 subjects. When the cue occurs, the subjects are supposed to implement movements or remain in a resting state. When the movement is executed, the movement onset can be located with hand the trajectory.

The movement trajectory $Hand_y$ is used to locate the movement onset of the 6 movement states. The 1-order difference of $Hand_y$ is taken, and then the 1-order Savitzky-Golay finite impulse response smoothing filter is used to smooth the signals. The following step differs when locating movement onset in these movement states.

For the two motions related to limb movement, \textit{elbow flexion} and \textit{elbow extension} lead to an increase in the amplitude of the hand trajectory. The trials whose variances are smaller than the threshold 0.05 are rejected as those containing invalid movement onset. By dividing the maximum value in the trajectory, the hand trajectories are normalized. The location where the normalized trajectory is larger than a threshold is regarded as the movement onset. In trials that contain heavy noise contamination, the movement onset cannot be located. These trials are manually removed. 

For the other four states, $Hand_y$ has a lower amplitude and is heavily influenced by noises. In these trials, trajectories are first normalized by dividing the maximal value of each trajectory. The function $f(x)=a*\exp{(-((x-b)/c)^2)+d})$ is used to fit the smoothed and normalized trajectories by parameter tuning on $a, b, c, d$. The symbol '$\exp$' denotes the exponential function. Trials that fulfil $a<0.05, c>100$ and $d>10$ are rejected. The movement onset is determined with a threshold criterion.

For the signals in the resting state, the amplitude of $Hand_y$ signals is supposed to be steady and has a small variance. The trials are rejected if the variances of trajectory are greater than the threshold 0.02. Because the trajectories in the resting state have no movement onsets. To facilitate the experiment, a fake movement onset is set to 2.5 $s$ after the beep sounds.

Because the EEG signals used in this work are located in the readiness potential section, the EEG signals in the readiness potential section are extracted after the localization of movement onset or fake movement onset. The data obtained are denoted as $\boldsymbol{X}^k\in \mathbb{R}^{N_c\times N_s \times N_t}$ or $X^k(t)\in \mathbb{R}^{N_c\times N_t}, t=1,...,N_s$. $N_c$ is EEG channel number. $N_s$ is the sample time of a trial. $N_t$ is the trial number. $k$ indicates whether the EEG tensor contains the data in the movement state or the resting state. Before the binary classification tasks, EEG signals are normalized by z-score normalization.

When evaluating STRCA and the proposed FBTRCA methods with this dataset, a 10-fold cross validation is applied. The dataset is divided into a training set and a testing set ten times. The classification performance is measured with a mean of 10 folds.

\subsection{Filter Bank Task-Related Component Analysis}
\subsubsection{Standard Task-Related Component Analysis}
\begin{figure}[htbp]
    \centering
    \includegraphics[width=\textwidth]{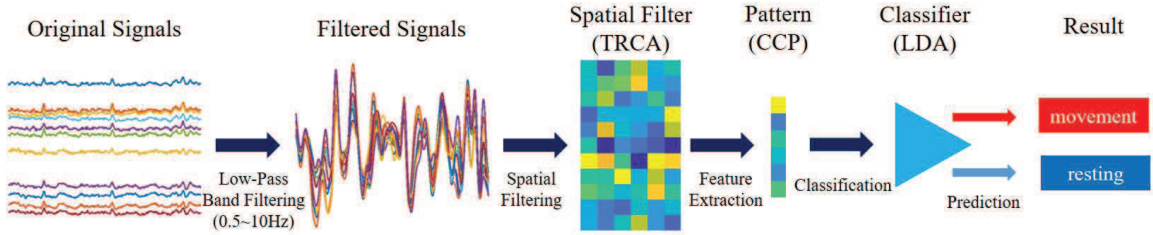}
    \caption{The structure of STRCA consists of the spatial filter TRCA and the extracted CCP features.}
    \label{fig:2.1}
\end{figure}

STRCA is used to classify the movement and resting state with MRCP signals in the readiness potential section \cite{duan_decoding_2021}. The STRCA consists of two components: (1) spatial filter TRCA and (2) CCP features. The extracted features are classified with the linear discriminated analysis (LDA) classifier. Figure \ref{fig:2.1} illustrates the structure of STRCA.

\paragraph{TRCA} The spatial filter of TRCA is designed by maximizing the reproducibility during the task \cite{nakanishi2018enhancing}.
In the multichannel EEG signals, the training set is supposed to be $X^k(t)\in\mathbb{R}^{N_c\times N_t}$. $k$ indicates that the class of this training set is movement or resting. Under the assumption of TRCA, $X(t)$ consists of two kinds of signals: (1) task-related signal $s(t)\in\mathbb{R}$ and (2) task-unrelated noise $n(t)\in\mathbb{R}$. The relationship between $X(t)$, $s(t)$ and $n(t)$ is expressed as:
\begin{align}
    X_{i,j}^k(t)
    =a_{1,i,j}^{k}s(t)+a_{2,i,j}^{k}n(t), i=1,...,N_c,j=1,...,N_t.
\end{align}
$y(t)$ is the linear sum of EEG signals $X(t)$, which is defined as:
\begin{align}
    y_{j}^{k}&(t)
    =\sum_{i=1}^{N_c}w_{i}^{k}X_{i,j}^{k}(t), j=1,...,N_t.
\end{align}
In TRCA, the task-related signal $s(t)$ is recovered from $y(t)$. The ideal solution is difficult to calculate but can be approached by maximizing the intertrial covariance. The covariance $C_{j_1,j_2}^{k}$ between the $j_1$-th trial and the $j_2$-th trial can be computed with:
\begin{align}
    C_{j_1,j_2}^{k}=Cov(y_{j_1}^{k}(t),y_{j_2}^{k}(t))
    =\sum_{i_1,i_2}^{N_c}w_{i_1}^k w_{i_2}^{k} Cov(X_{i_1,j_1}^{k} (t),X_{i_2,j_2}^{k} (t)).
\end{align}
The covariances of all trials are summed to obtain the combination of all trials:
\begin{align}
  \sum_{\substack{j_1,j_2=1 \\ j_1\not= j_2}}^{N_t}C&_{j_1,j_2}^{k}
    =\sum_{\substack{j_1,j_2=1 \\ j_1\not= j_2}}^{N_t}Cov(y_{j_1}^{k}(t),y_{j_2}^{k}(t))\notag\\
    &=\sum_{\substack{j_1,j_2=1 \\ j_1\not= j_2}}^{N_t}\sum_{i_1,i_2=1}^{N_c}w_{i_1}^{k}w_{i_2}^{k}Cov(X_{i_1,j_1}^{k}(t),X_{i_2,j_2}^{k}(t))=\boldsymbol{w}^{T}S^k\boldsymbol{w}.
\end{align}
To avoid infinite solutions of $\boldsymbol{w}$, the variance of $y_{j}^{k}(t)$ is constrained to 1:
\begin{align}
  \sum_{j_1,j_2=1}^{N_t}C&_{j_1,j_2}^{k}=\sum_{j_1,j_2=1}^{N_t}Cov(y_{j_1}^{k}(t),y_{j_2}^{k}(t))\notag\\
  &=\sum_{j_1,j_2=1}^{N_t}\sum_{i_1,i_2=1}^{N_c}w_{i_1}^{k}w_{i_2}^{k}Cov(X_{i_1,j_1}^{k}(t),X_{i_2,j_2}^{k}(t))=\boldsymbol{w}^{T}Q^{k}\boldsymbol{w}.
\end{align}
The constrained spatial filter can be obtained by maximizing the generalized eigenvalue equation $J$. The expression of $J$ is
\begin{equation}
    J=\frac{\boldsymbol{w}^{T}S^k\boldsymbol{w}}{\boldsymbol{w}^{T}Q^{k}\boldsymbol{w}}.
\end{equation}
After solving the generalized eigenvalue problem, the eigenvectors with maximum eigenvalues are selected as the eigenvectors used in the spatial filter. Three eigenvectors are adopted in TRCA. These eigenvectors from two classes are then combined into the TRCA spatial filter. The TRCA spatial filter we obtained are $W\in \mathbb{R}^{N_c\times 6}$ for movement onset detection. 

\paragraph{CCP} With the training set of EEG data, $\boldsymbol{X}^{k}\in\mathbb{R}^{N_c\times N_s\times N_t},k=1,2$, we can obtain the CCP templates $\hat{X}^{k}=\sum_{j=1}^{N_t}\boldsymbol{X}^{k}/N_t \in \mathbb{R}^{N_c\times N_s},k=1,2$ for each of two classes. 
The EEG signal of the trial from which we aim to extract features is $X\in \mathbb{R}^{N_c\times N_s}$. 
Given the TRCA spatial filter $W$, we extract the CCP after the EEG signals are transformed with $W$.
Three kinds of correlation coefficients are considered in STRCA:

\noindent (1) Correlation coefficients between filtered signals: 
\begin{eqnarray}
    &X_k=\hat{X}^{k};X_*=X;\\
    &\rho_{1,k}=corr(X_*^TW, X_k^TW),k=1,2;
    \label{equ:coef1}
\end{eqnarray}

\noindent (2) Correlation coefficients between filtered signals with a canonical correlation analysis projection:
\begin{eqnarray}
    &X_k=\hat{X}^{k};X_*=X;\\
    &[A_k, B_k]=cca(X_*^TW, X_k^TW)\\
    &\rho_{2,k}=corr(X_*^TWB_k, X_k^TWB_k),k=1,2;
    \label{equ:coef2}
\end{eqnarray}

\noindent (3) Correlation coefficients between the distances of filtered signals:
\begin{eqnarray}
    &X_k=\hat{X}^{k}-\hat{X}^{3-k}; X_*=X-\hat{X}^{3-k};\\
    &[A_k, B_k]=cca(X_*^TW, X_k^TW)\\
    &\rho_{3,k}=corr(X_*^TWB_k, X_k^TWB_k),k=1,2.
    \label{equ:coef3}
\end{eqnarray}
In the above equations, canonical correlation analysis is used to optimize the correlation between the templates $\hat{X}^{k}$ and $X$. The function symbols $corr$ and $cca$ indicate the calculation process of the correlation coefficient and the process of CCA analysis, respectively.

For each trial, we can obtain six features. These six features are called CCP features in the following section.

\subsubsection{Filter Bank TRCA}
\begin{figure}[htbp]
    \centering
    \includegraphics[width=\textwidth]{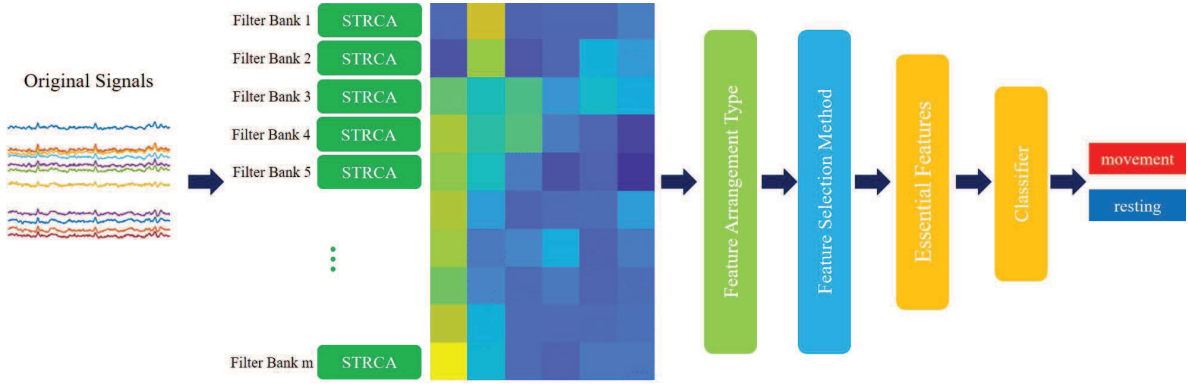}
    \caption{The structure of FBTRCA. CCP features are extracted from EEG signals in various filter banks, and a total of $6\times m$ features are obtained. Feature selection methods are used to extract the essential features in these features. A binary classifier is used to classify the selected essential features and predict the state of EEG signals (movement or resting).}
    \label{fig:2.2}
\end{figure}

This study proposes an FBTRCA method to enhance pattern decoding in MRCP analysis. Figure \ref{fig:2.2} shows the flowcharts of the proposed FBTRCA. The FBTRCA method consists of three major procedures: (1) filter bank analysis, (2) CCP feature extraction and (3) feature selection.

First, in the filter bank technique, the subbands are decomposed with multiple filters that have different pass-bands. In this study, the bandpass filters for extracting subband components from original EEG signals are 8-order infinite impulse Butterworth filter.

Second, after the filter bank analysis, STRCA is applied to each of the subbands separately, resulting in six CCP features for each subband. If the number of subbands is denoted as $m$, the number of CCP features extracted from all subbands is 6$\times m$=$6m$. The essential features are extracted from the $6m$ features in all subbands with one of the feature arrangement types. The feature arrangement type refers to the arrangement of $6m$ CCP features when the feature selection method is applied. 

Third, the selected essential features are classified with the a binary classifier. Three classifiers are considered and compared in this study, including LDA, neural network (NN) and support vector machine (SVM).

\subsubsection{Frequency Range Settings}
In the decomposition of subbands, the decomposed EEG signals and classification accuracies differ with different frequency range settings of filters. In the MI analysis, the frequency range of filter banks was equipped with equally spaced band widths in alpha and beta rhythms, e.g., 4$\sim$8 $Hz$, 8$\sim$12 $Hz$, ..., 36$\sim$40 $Hz$ \cite{zhang_temporally_2019}. In the SSVEP analysis, the frequency range of filter banks started at $n\times$8 $Hz$ and ended at a fixed frequency, e.g., 8$\sim$88 $Hz$, 16$\sim$88 $Hz$, ..., 80$\sim$88 $Hz$, $n=1,2,...,10$ \cite{Chen_2015}.

In MRCP analysis, the frequency range setting is undetermined. Considering that MRCP signals are a kind of EEG signal with low frequencies, the maximum of frequency range is set to 10 $Hz$. Three frequency arrangement settings are compared, including $M_1$, $M_2$ and $M_3$.

\begin{figure}[htbp]
    \centering
    \subfigure[$M_1$]{\includegraphics[width=0.323\textwidth]{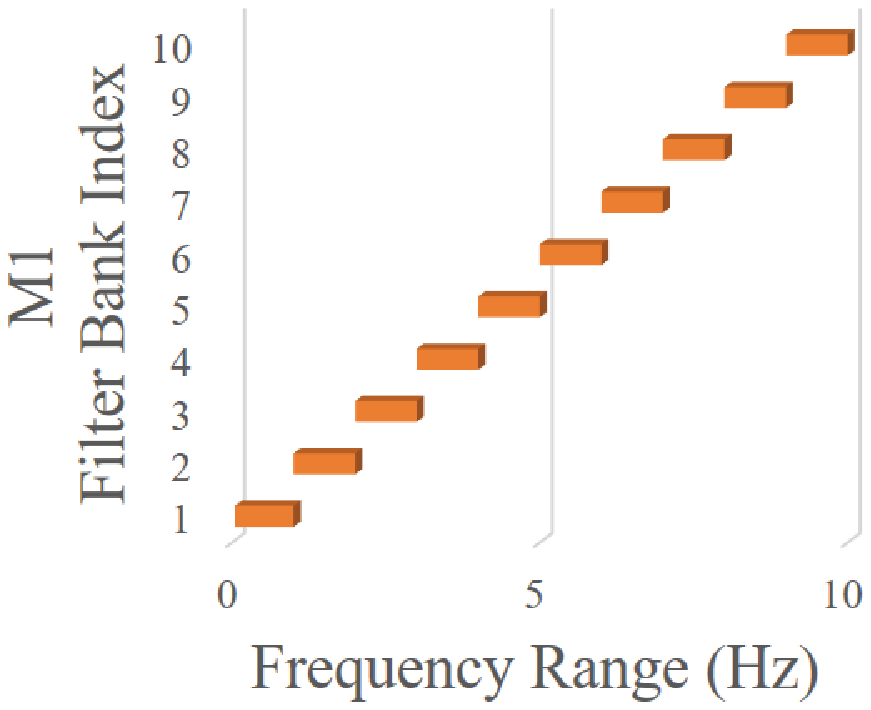}\label{fig:2.3a}}
    \subfigure[$M_2$]{\includegraphics[width=0.323\textwidth]{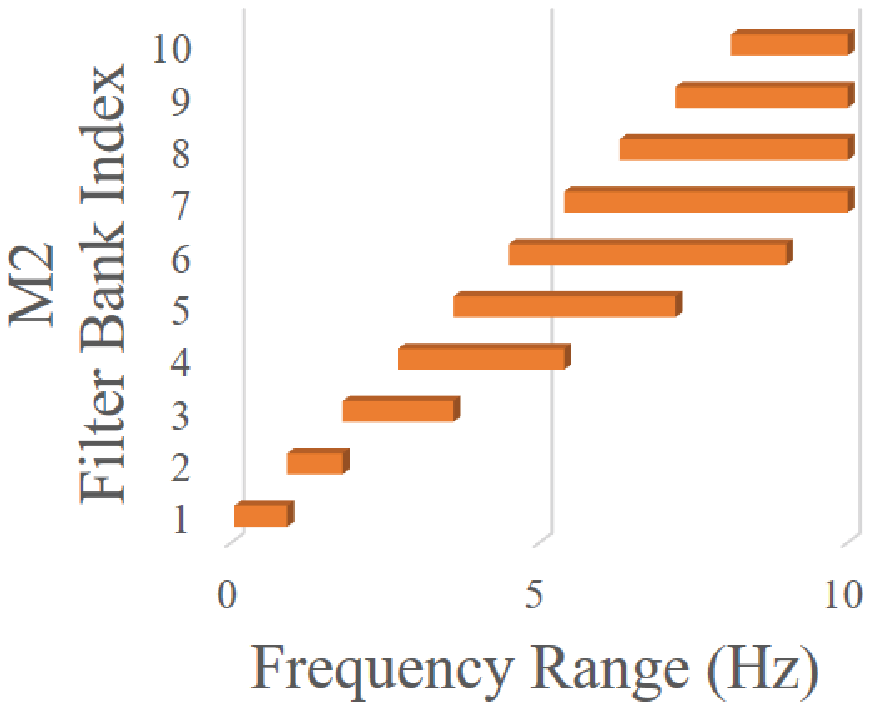}\label{fig:2.3b}}
    \subfigure[$M_3$]{\includegraphics[width=0.323\textwidth]{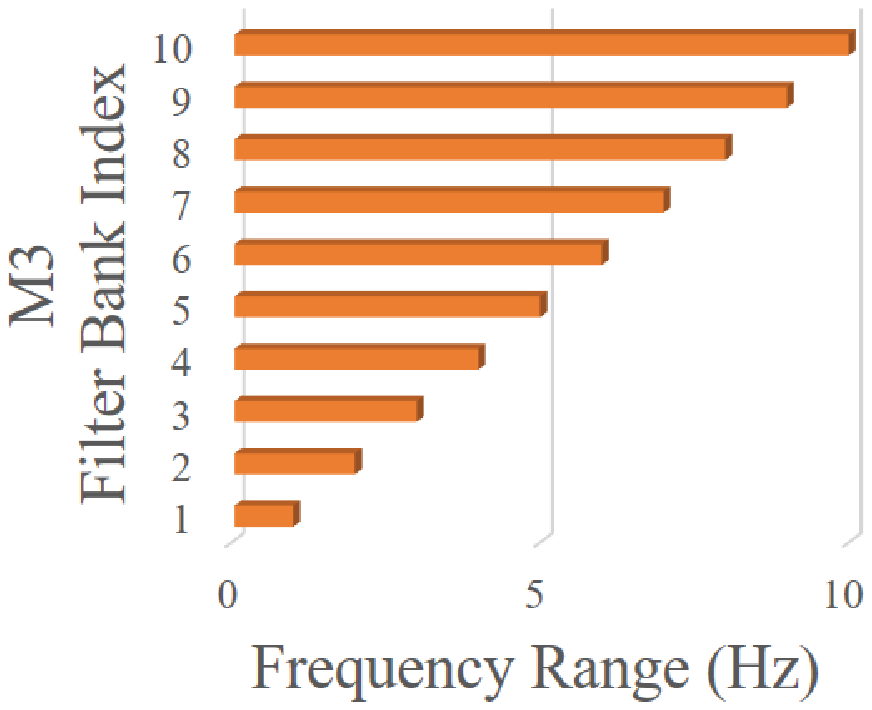}\label{fig:2.3c}}
    \caption{Frequency range settings of subbands for filter bank design. $M_1$: subbands with equally spaced bandwidths (e.g., 0.05$\sim$1 $Hz$, 1$\sim$2 $Hz$, ..., 9$\sim$10 $Hz$).; $M_2$: subbands whose stop frequency is twice as high as the start frequency (e.g., 0.05$\sim$0.9 $Hz$, 0.9$\sim$1.8 $Hz$, 1.8$\sim$3.6 $Hz$, ..., 8.1$\sim$10 $Hz$).; $M_3$: subbands that start at a fixed frequency (e.g., 0.05$\sim$1 $Hz$, 0.05$\sim$2 $Hz$, ..., 0.05$\sim$10 $Hz$). Considering that the MRCP is EEG signals with low frequencies, the maximum frequency of the range is set to $10 Hz$. In this figure, the number of filter banks is 10 ($m$=10).}
    \label{fig:2.3}
\end{figure}

\noindent(1) $M_1$ Figure \ref{fig:2.3a}: The frequency range setting in $M_1$ is similar to that in FBCSP but with a different maximum and minimum frequencies. The subbands in $M_1$ are equipped with equal-space band widths.

\noindent(2) $M_2$ Figure \ref{fig:2.3b}: The frequency range setting in $M_2$ corresponds to the harmonic frequency bands. The high cut-off frequency is twice as high as the low cut-off frequency.

\noindent(3) $M_3$ Figure \ref{fig:2.3c}: The frequency range setting in $M_3$ is similar to the best setting in FBCCA. One of the two ends of the subbands is a fixed value. Because the MRCP signals are in a low-frequency band, the low cut-off is fixed.

\subsubsection{Feature Arrangement Types}
\begin{figure}[htbp]
    \centering
    \includegraphics[width=\textwidth]{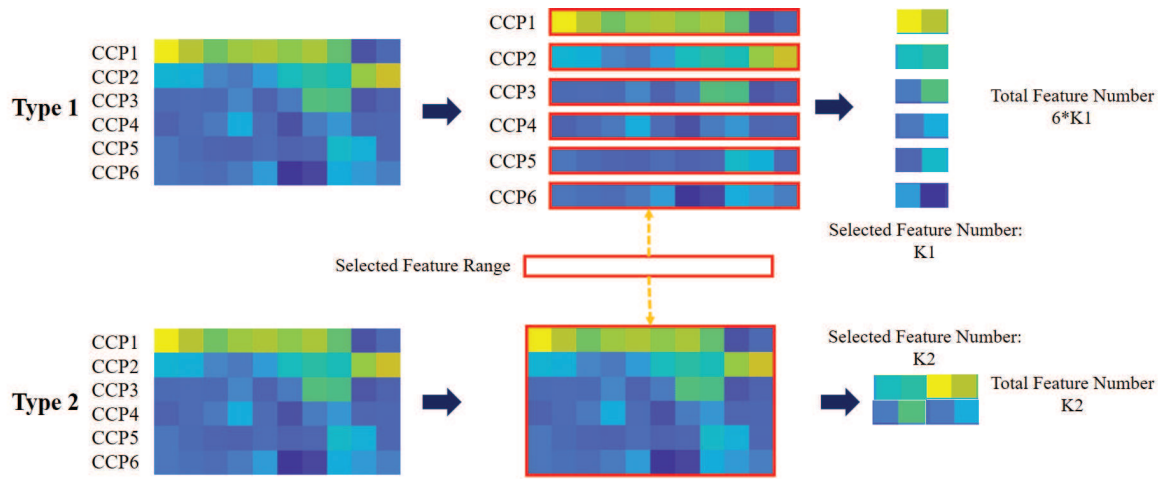}
    \caption{There are two arrangements when selecting the features across these filter banks. Type 1: A feature selection method is applied to each feature in CCP respectively. Type 2: A feature selection method is applied to all six features in CCP simultaneously.}
    \label{fig:2.4}
\end{figure}

From each of all subbands, six CCP features are extracted. The number of all features is 6$\times$m=$6m$. When selecting essential features with feature selection methods from $6m$ features, there are two feature arrangement types (Figure \ref{fig:2.4}).

\noindent Type 1: The feature selection method is applied to each feature in the CCP individually. The feature selection method selects K1 essential features from $m$ features. The feature selection method is applied six times. Finally, 6*K1 essential features are selected with feature selection methods in total. 

\noindent Type 2: The feature selection method is applied to all six features in CCP simultaneously. The feature selection method is applied only once and totally K2 essential features are selected from $6m$ features.

\subsubsection{Feature Selection Methods}
Mutual-information based approaches are an important feature selection paradigm in data mining. In the FBCSP method of MI analysis, feature selection based on mutual information plays a significant role in optimizing the CSP features in all subbands. In this study, eight mutual-information based feature selection methods are compared to find a suitable one in selecting CCP features of pre-movement decoding. The compared feature selection methods include:

\noindent (1) {Mutual Information Quotient (MIQ)\cite{ding2003miq}}

\noindent (2) {Maximum Relevance (MAXREL)\cite{peng2005mrmr}}

\noindent (3) {Minimum Redundancy (MINRED)\cite{peng2005mrmr}}

\noindent (4) {Minimum Redundancy Maximum Relevance (MRMR)\cite{peng2005mrmr}}

\noindent (5) {Quadratic Programming Feature Selection (QPFS)\cite{lujan2010qpfs}}

\noindent (6) {Conditional Infomax Feature Extraction (CIFE)\cite{lin2006cife}}

\noindent (7) {Conditional Mutual Information Minimization (CMIM)\cite{herman2013cmim}}

\noindent (8) {Maximum Relevance Minimum Total Redundancy (MRMTR)\cite{nguyen2014mrmtr}}

\subsubsection{Binary Classifiers}
In STRCA, three binary classifiers have been compared, including linear discriminate analysis (LDA), support vector machine (SVM) and neural network (NN). Because the feature number of STRCA is a fixed value, the simple LDA classifier shows the best performance among the three classifiers. In the proposed FBTRCA, there are more than six features, and the feature numbers change due to the feature selection settings. Therefore, it is necessary for the three classifiers to be compared as well.

\subsection{Cross Validation and Testing in SSSF} 

In SSSF, a cross validation and testing (CVT) method is used to find the frequency range in which the extracted features have the best classification performance. Here the CVT is used as a benchmark method. In the validation of CVT, the training set is regarded as a new dataset. A 9-fold cross validation is applied to the new dataset. In the new dataset, i.e., the training set, all labels of trials are known and the best frequency range can be found with the 9-fold cross validation. In each fold of the 10-fold cross validation, a 9-fold cross validation is applied to find the best frequency range. The classification performance of CVT is measured with the testing set of 10-fold cross validation. The CVT is actually STRCA which is applied to EEG signals in the best-fit frequency range. The binary classifier in CVT is set to the LDA classifier, which is the same as that in STRCA.

\section{Result}
\label{sec:resul}

There were six movement states and a resting state for 15 subjects in this study. The STRCA or FBTRCA is used to classify two states (movement vs resting) with 10-fold cross validation. Therefore, 6 (movement states) $\times$ 15 (subjects) $\times$ 10 (folds) accuracies are obtained. The mean of these accuracies is taken to evaluate the classification performance.

The purpose of this study is to incorporate the filter bank technique into STRCA and propose a new FBTRCA. Two steps are necessary to achieve this goal:

\noindent (1) decide on the frequency range settings;

\noindent (2) evaluate the parameters K1 and K2 in two feature arrangement types.

In the first step, the properties of each filter bank will be determined, including the number of filter banks and the frequency range of each filter bank. In the second step, K1 and K2 for each of mutual-information based feature selection methods will be evaluated. Three binary classifiers are also compared in the second step.

\subsection{Analysis on Frequency Range Settings}
\label{ssec:frs}
\begin{figure}[htbp]
    \centering
    \subfigure[$M_1$]{\includegraphics[width=0.323\textwidth]{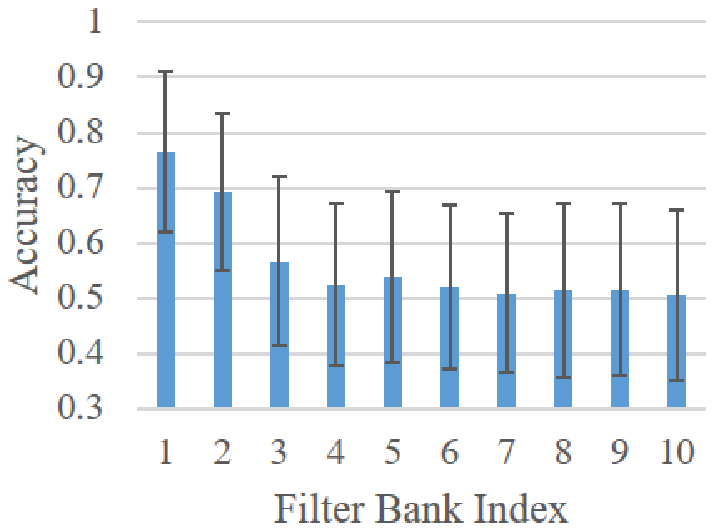}}
    \subfigure[$M_2$]{\includegraphics[width=0.323\textwidth]{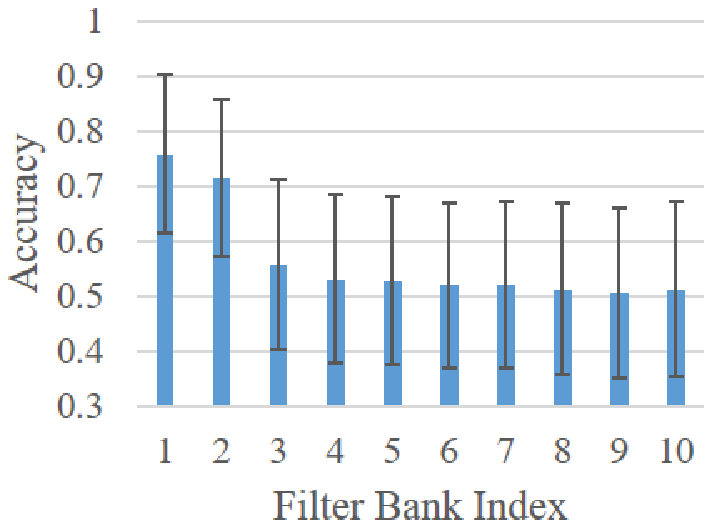}}
    \subfigure[$M_3$]{\includegraphics[width=0.323\textwidth]{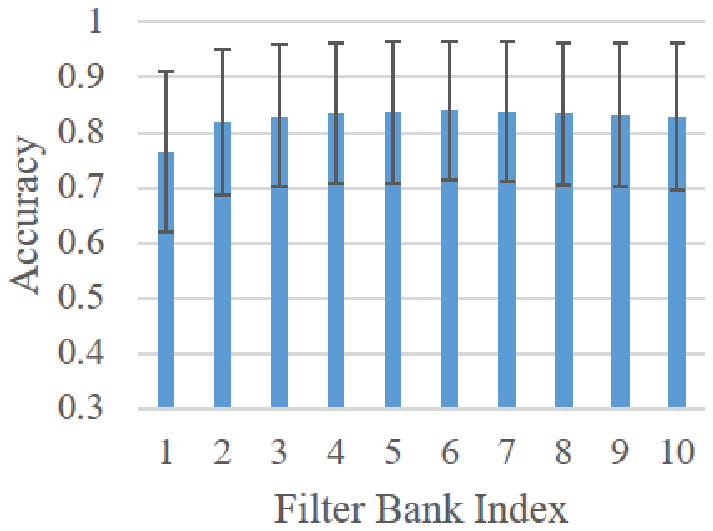}}
    \caption{Classification accuracies of STRCA in three frequency range settings. In both settings $M_1$ and $M_2$, the accuracies decrease as the filter bank index increases. This means that STRCA cannot tell the differences between movement state and resting state. In the setting $M_3$, the accuracies remain stable with an acceptable accuracy range. This means that $M_3$ is an acceptable setting among the three frequency arrangement settings.}
    \label{fig:3.1}
\end{figure}

In the analysis on frequency range settings, STRCA is applied to the filter banks of three settings: $M_1$, $M_2$ and $M_3$. Figure \ref{fig:3.1} shows the classification accuracies of each filter bank in three settings. 

In setting $M_1$, the accuracy decreases to 0.5 as the filter bank index increases. The accuracy of the $M_2$ setting has the same trend as that in the $M_1$ setting. This means that STRCA fails to classify the movement state and resting state in subbands without low frequencies. The two frequency range settings are not suitable for the combination of STRCA and the filter bank technique. In setting $M_3$, the accuracies in these filter banks are acceptable, and STRCA successfully classifies the movement state and resting state.

The main difference between $M_3$ and $M_1$ or $M_2$ is that the frequency ranges of filter banks in $M_3$ cover the subbands at low frequencies. The subbands at low frequencies maintain the main information necessary for STRCA.

In the design of filter banks of the proposed FBTRCA, the frequency range setting $M_3$ is adopted with modifications. The low cut-off frequency is shifted slightly from 0.5 $Hz$ to 0.05 $Hz$ with step 0.05 $Hz$, and the high cut-off frequency remains the same as that in $M3$. The adopted subbands number is 10$\times$10 in total.

\begin{figure}[htbp]
    \centering
    \subfigure[]{\includegraphics[width=0.24\textwidth]{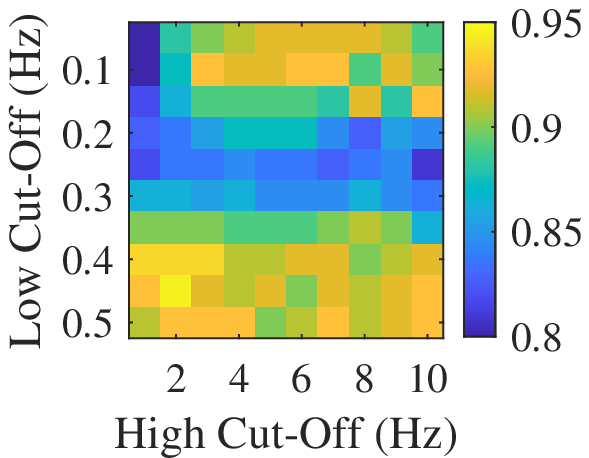}}
    \subfigure[]{\includegraphics[width=0.24\textwidth]{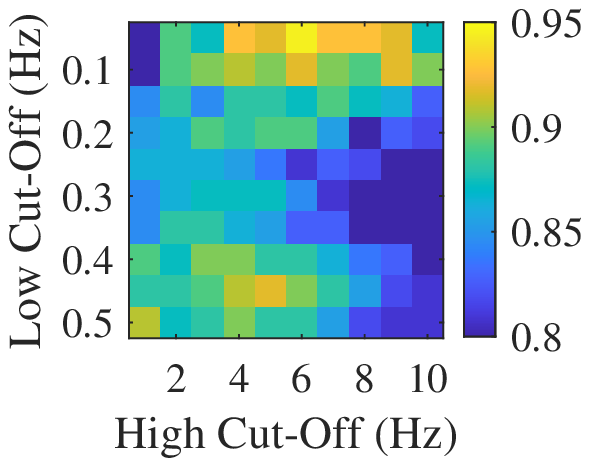}}
    \subfigure[]{\includegraphics[width=0.24\textwidth]{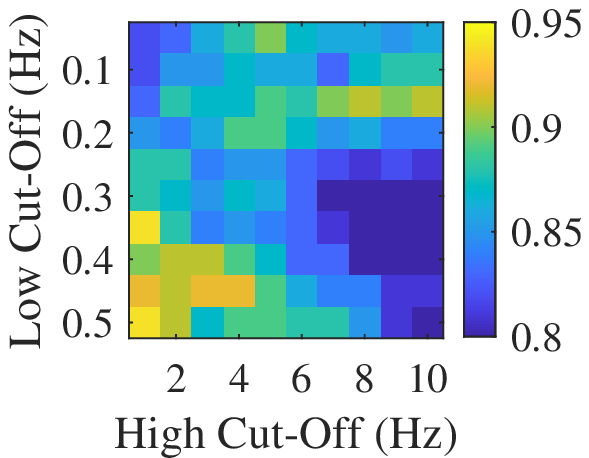}}
    \subfigure[]{\includegraphics[width=0.24\textwidth]{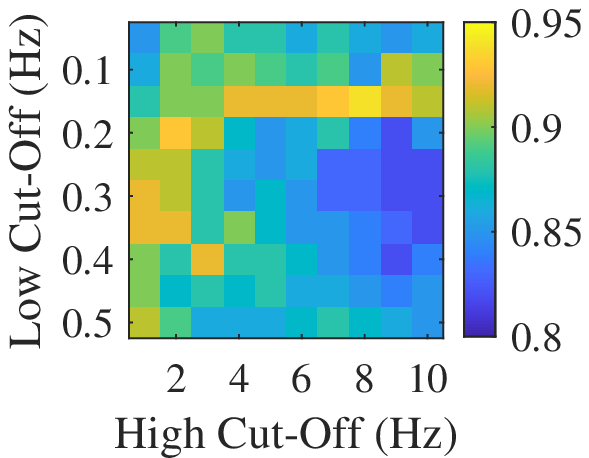}}
    \caption{Average classification accuracies of STRCA in different subbands. (a) Subject 1, \textit{elbow flexion}; (b) Subject 1, \textit{elbow extension}; (c) Subject 2, \textit{elbow flexion}; (d) Subject 2, \textit{elbow extension}. For different subjects or movement states, the subband that has the best classification performance differs. This is the reason why it is necessary for STRCA to incorporate the filter bank technique and develop the proposed FBTRCA method.}
    \label{fig:3.2}
\end{figure}

In Figure \ref{fig:3.2}, the classification accuracies of STRCA in these subbands are given. For different subjects and movement states, the subbands with the highest accuracy are different. It is hard to decide a suitable subband for STRCA. FBTRCA is proposed to solve this problem.

\subsection{Analysis on Feature Selection Methods}

\paragraph{Classification Accuracy}

\begin{figure}[htbp]
    \centering
    \subfigure[Parameter searching on K1 (left) and K2 (right), classified with LDA]{\includegraphics[width=0.57\textwidth]{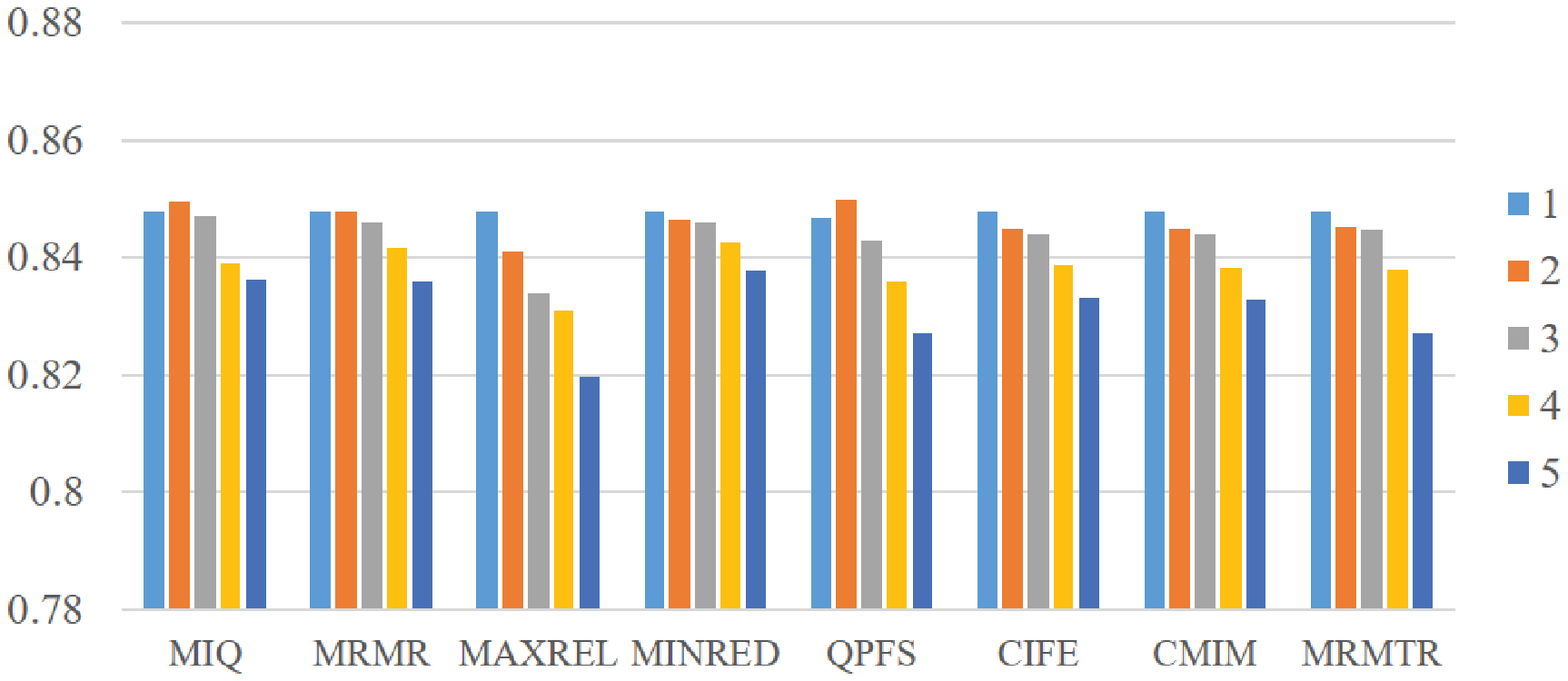}
    \includegraphics[width=0.38\textwidth]{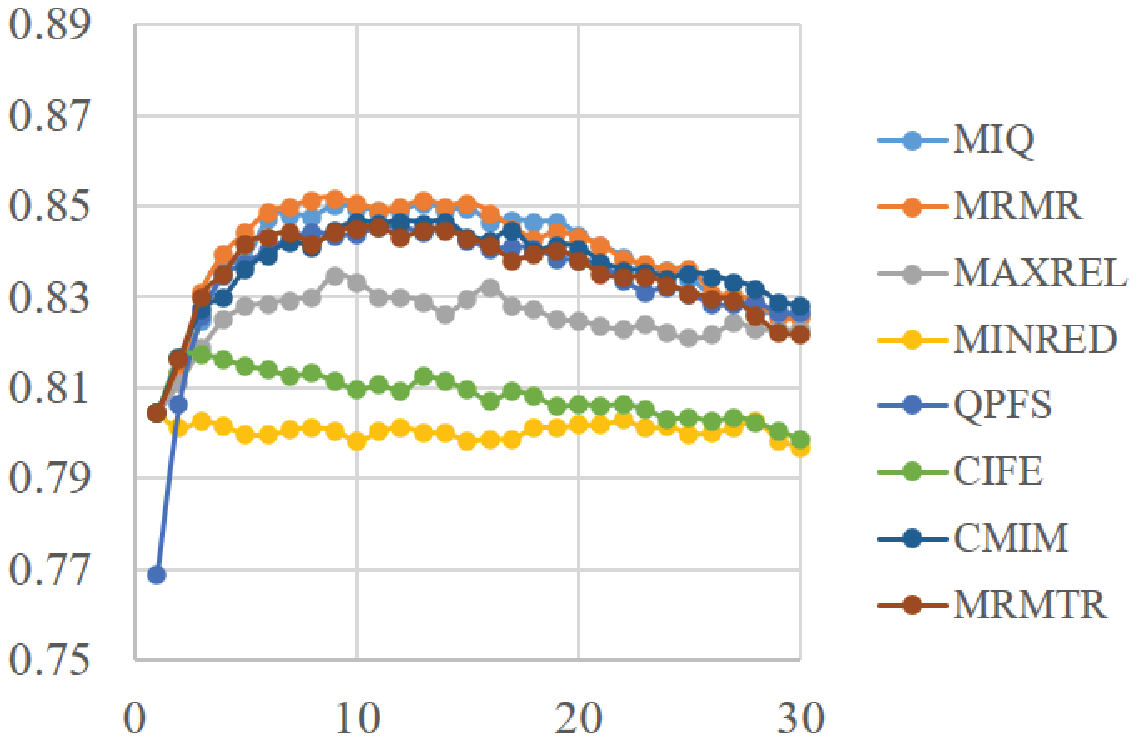}
    \label{fig:3.3a}}
    \subfigure[Parameter searching on K1 (left) and K2 (right), classified with SVM]{\includegraphics[width=0.57\textwidth]{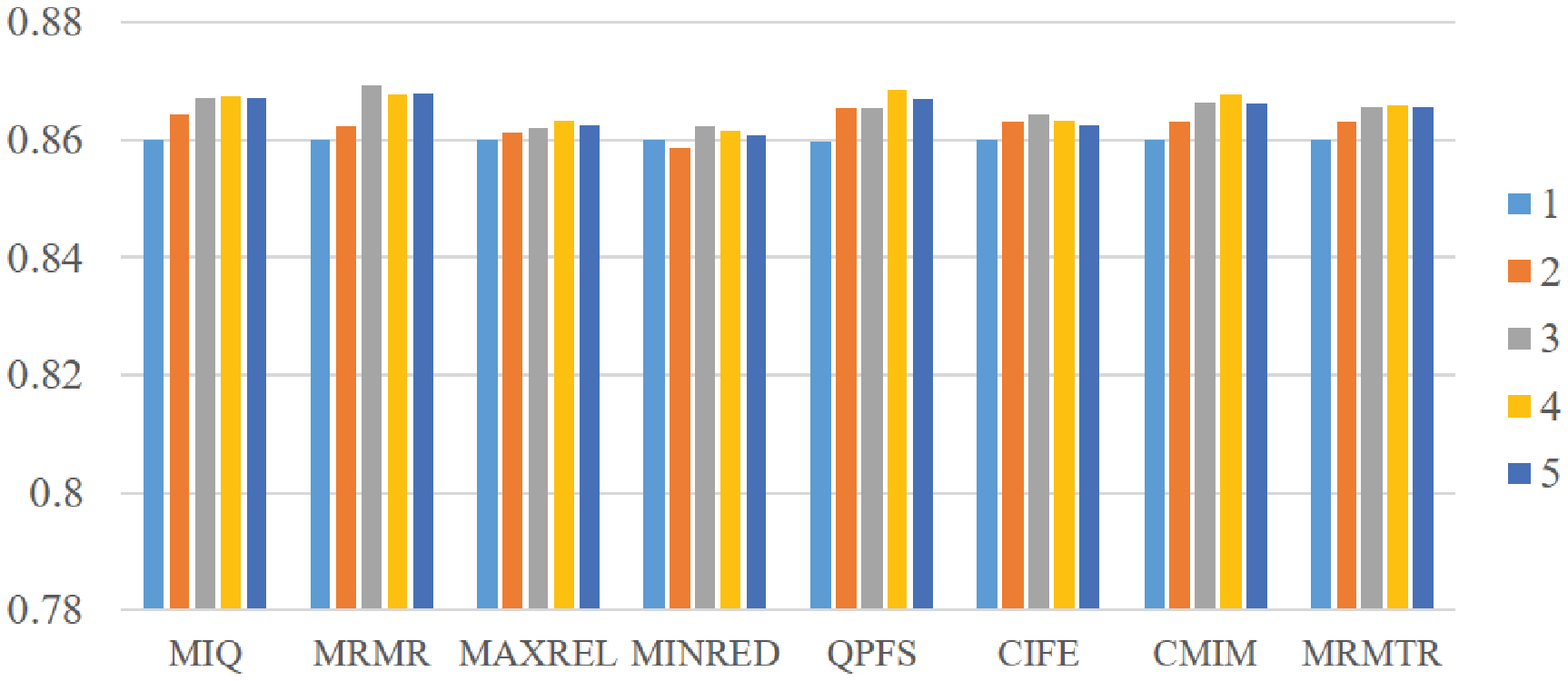}
    \includegraphics[width=0.38\textwidth]{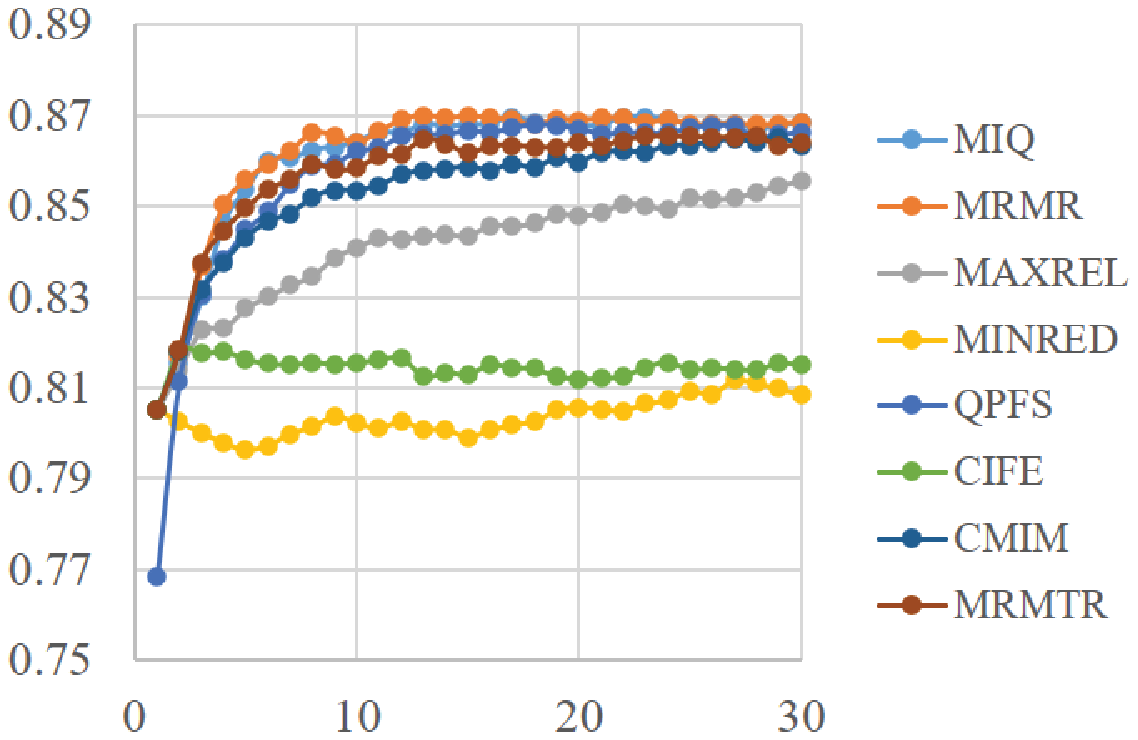}
    \label{fig:3.3b}}
    \subfigure[Parameter searching on K1 (left) and K2 (right), classified with NN]{\includegraphics[width=0.57\textwidth]{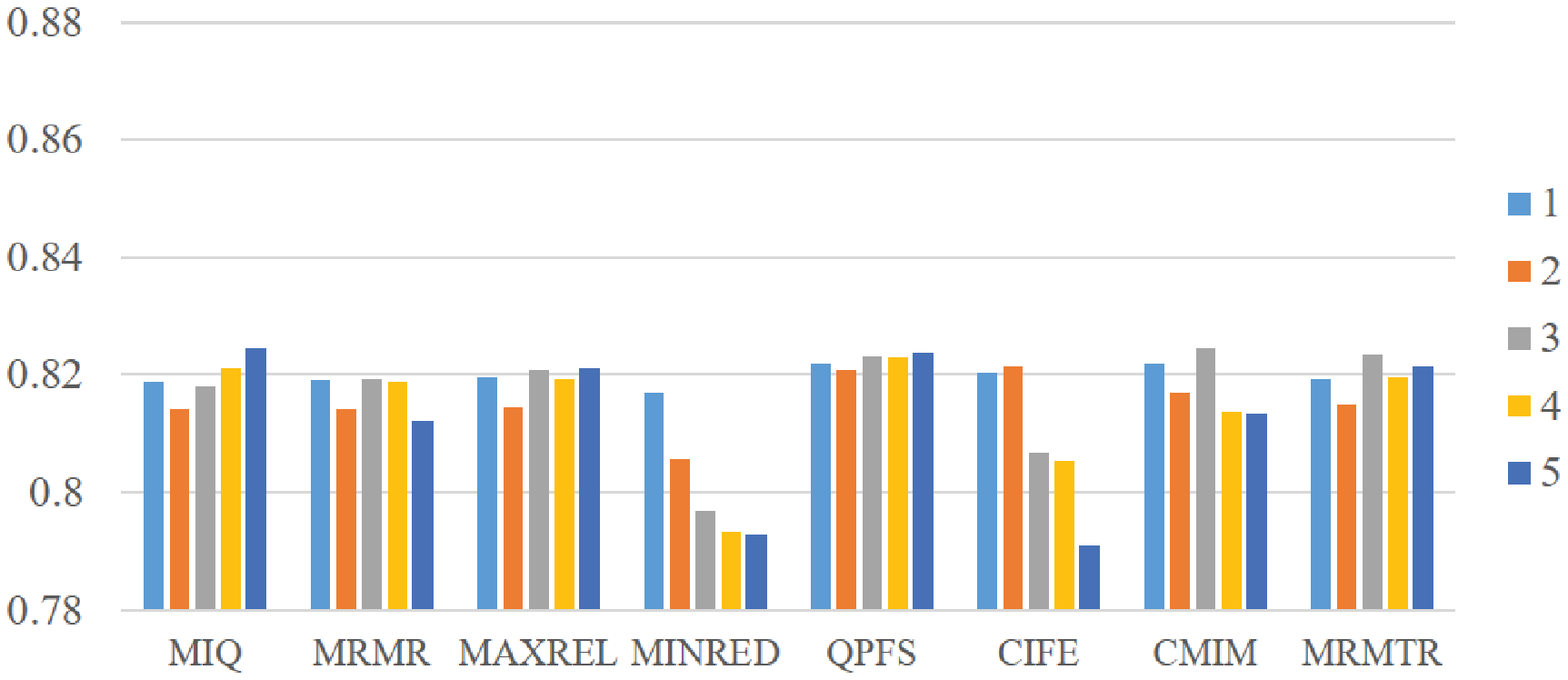}
    \includegraphics[width=0.38\textwidth]{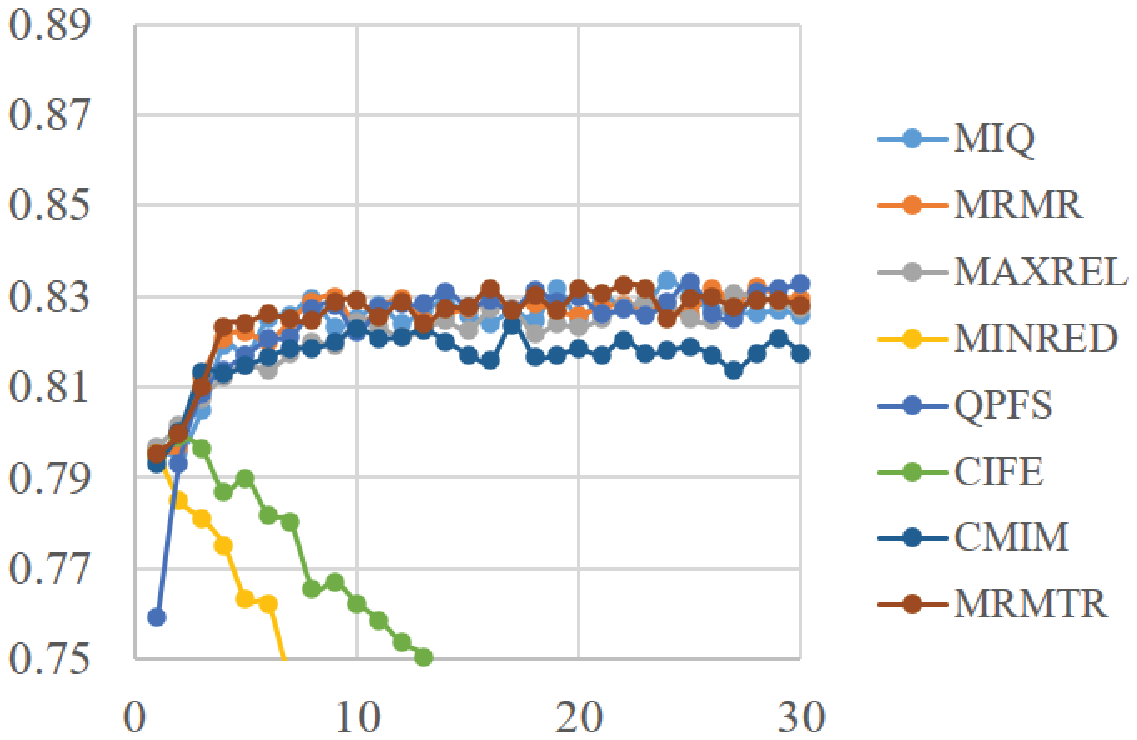}
    \label{fig:3.3c}}
    \caption{Parameter searching of K1 and K2 in two feature arrangement types, K1$=$1,2,3,4,5 (left) and K2$=$1,2,3,...,30 (right). The selected essential features with feature selection methods are classified with (a) LDA classifier, (b) SVM classifier and (c) NN classifier with hidden neural number 10, respectively. The purpose of this figure is to (1) select the best K1 and K2 for two feature arrangement types and (2) find the best feature selection method among eight feature selection methods and (3) decide to use which classifier. The standard deviation is omitted in this figure for better comparison and easier selection.}
    \label{fig:3.3}
\end{figure}

After applying STRCA to 100 subbands, 6$\times$100 CCP features are extracted. Later, feature selection methods are used to select the essential features with certain feature arrangement type. The essential features are classified with binary classifiers. In Figure \ref{fig:3.3}, the classification performances of eight mutual-information based feature selection methods on two feature arrangement types are compared. The essential features are classified with LDA classifier, SVM classifier (linear kernel) and NN classifier (hidden neuron number: 10) respectively.

The results in Figure \ref{fig:3.3} are analysed from three perspectives: binary classifiers, parameter searching on K1 and parameter searching on K2. 

\noindent (1) \textit{Classifiers} In Figure \ref{fig:3.3a}, the maximum accuracies of searching K1 (left) and K2 (right) are approximately 0.8400 and 0.8500, respectively. In Figure \ref{fig:3.3b}, the two accuracies are approximately 0.8600 and 0.8700. In Figure \ref{fig:3.3c}, the two accuracies are approximately 0.8200 and 0.8300. Hence the SVM classifier is more suitable for the proposed FBTRCA than the other two binary classifiers. In Figure \ref{fig:3.3a}, the accuracy has a decreasing trend as K2 increases. The possible reason is that the number of features is too large for the LDA classifier to deal with.

\noindent (2) \textit{K1 Parameter} In Figure \ref{fig:3.3b} (left), K1 increases from 1 to 5 and the total feature number increases from 6 to 30. The accuracies have an increasing trend as K1 increases. In Table \ref{tab:3.1}, the best K1 value for each feature selection method is listed.

\begin{table}[htbp]
    \caption{Parameter K1 of Feature Selection Methods}
    \centering
    \footnotesize
    \begin{tabular}{c|c|c|c|c|c|c|c|c}
        \toprule
        Method & MIQ & MRMR & MAXREL & MINRED & QPFS & CIFE & CMIM & MRMTR\\
        \midrule
        K1 Value&4&3&4&3&4&3&4&4\\
        \bottomrule
    \end{tabular}
    \label{tab:3.1}
\end{table}

\noindent (3) \textit{K2 Parameter} In Figure \ref{fig:3.3b} (right), K2 increases from 1 to 30. The accuracy trajectories become steady when K2 is larger than 13. 

\begin{table}[htbp]
    \caption{Parameter K2 of Feature Selection Methods}
    \centering
    \footnotesize
    \begin{tabular}{c|c|c|c|c|c|c|c|c}
        \toprule
        Method & MIQ & MRMR & MAXREL & MINRED & QPFS & CIFE & CMIM & MRMTR\\
        \midrule
        K2 Value&17&13&30&27&18&2&29&28\\
        \bottomrule
    \end{tabular}
    \label{tab:3.2}
\end{table}

With the parameters in Table \ref{tab:3.1} and Table \ref{tab:3.2}, the best classification accuracy for each feature selection method can be found. Table \ref{tab:3.3} gives the best accuracies of these methods. Because the accuracies of MAXREL, MINRED and CIFE are unsatisfactory, the results of the three methods are not included in Table \ref{tab:3.3}.

\begin{table}[htbp]
    \caption{Best Accuracies of Feature Selection Methods}
    \centering
    \footnotesize
    \begin{tabular}{c|c|c|c|c|c|c}
        \toprule
        \multicolumn{2}{c|}{Method}& MIQ & MRMR & QPFS & CMIM & MRMTR\\
        \midrule
        \multirow{2}{*}{Accuracy}& Type1 & 0.8674$\pm$0.1011 & 0.8691$\pm$0.1005 & 0.8685$\pm$0.1040 & 0.8676$\pm$0.1047 & 0.8660$\pm$0.1052\\
        \cmidrule{2-7}
        & Type2 & 0.8697$\pm$0.1010 & 0.8700$\pm$0.1022 & 0.8682$\pm$0.1021 & 0.8650$\pm$0.1017 & 0.8656$\pm$0.1026\\
        \bottomrule
    \end{tabular}
    \label{tab:3.3}
\end{table}

\paragraph{Time Consumption}
\begin{figure}[htbp]
    \centering
    \includegraphics[width=0.8\textwidth]{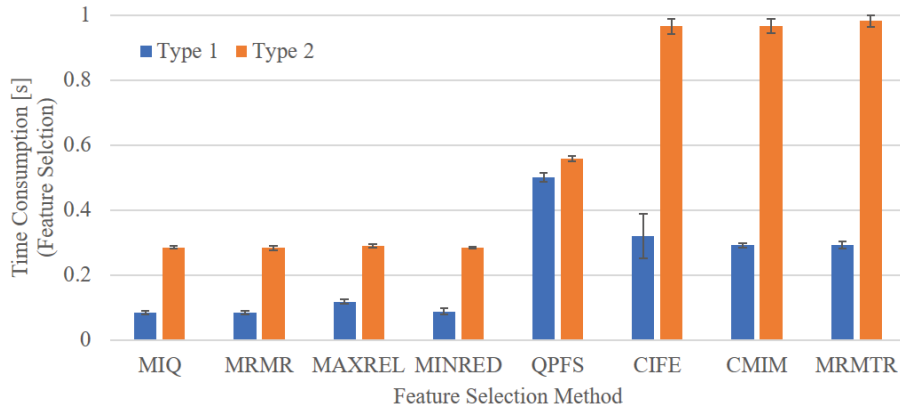}
    \caption{Time consumption of feature selection methods. Considering that the consumed time when applying STRCA to each subband are the same for these feature selection methods, only the consumed time during feature selection is considered.}
    \label{fig:3.5}
\end{figure}

The classification accuracies are close to each other. Therefore, the time consumption of these methods is calculated. Considering that the steps before feature selection are the same, only the consumed time during selecting features are calculated. Figure \ref{fig:3.5} gives the time consumption for feature selection. In Figure \ref{fig:3.5}, MIQ and MRMR are the two methods that consume less time.

\subsection{Comparison against Benchmarks}

Considering the classification accuracy and time consumption simultaneously, MRMR is used as the feature selection method in the proposed FBTRCA method. In addition, the feature arrangement type 2 is preferred in FBTRCA. The parameter K2 in feature arrangement type 2 is set to 13 because the accuracies remain steady as K2 becomes larger than 13 in Figure \ref{fig:3.3b}.

To show the state of the art of the proposed FBTRCA method, comparisons against the previous methods are implemented. These methods are listed in Table \ref{tab:3.4}. In Table \ref{tab:3.4}, the classifiers and the maximum and minimum frequency ranges are given. Bank number refers to the filter bank numbers used in that method. The STRCA number refers to how many STRCAs are applied to train this method. The STRCA number of STRCA is definitely 1. In the proposed FBTRCA, the STRCA number equals the filter bank number because multiple STRCAs are applied to extract features from the subbands. In the CVT method, the selection of the best subband is fulfilled with a 9-fold cross validation. Therefore, the STRCA number equals the quotient of the bank number times 9.

\begin{table}[htbp]
    \centering
    \caption{Information about the compared benchmarks}
    \footnotesize
    \begin{tabular}{c|c|c|c|c|c}
        \toprule
        Name & Classifier & Frequency Range & Bank Number & STRCA Number & Reference \\
        \midrule
        STRCA1 & LDA &  0.5$\sim$10 $Hz$ & 1 & 1 & \cite{duan_decoding_2021} \\
        \midrule
        STRCA2 & LDA & 0.05$\sim$10 $Hz$ & 1 & 1 & \cite{duan_decoding_2021} \\
        \midrule
        CVT      & LDA & 0.05$\sim$10 $Hz$ & 100 & 100$\times$9 & \cite{duan_decoding_2021,jeong_decoding_2020}\\
        \midrule
        FBTRCA(NN)   & NN  & 0.05$\sim$10 $Hz$ & 100 & 100 & \textit{proposed}\\
        \midrule
        FBTRCA(LDA)   & LDA & 0.05$\sim$10 $Hz$ & 100 & 100 & \textit{proposed}\\
        \midrule
        FBTRCA(SVM)   & SVM & 0.05$\sim$10 $Hz$ & 100 & 100 & \textit{proposed}\\
        \bottomrule
    \end{tabular}
    \label{tab:3.4}
\end{table}

\begin{figure}[htbp]
    \centering
    \includegraphics[width=0.8\textwidth]{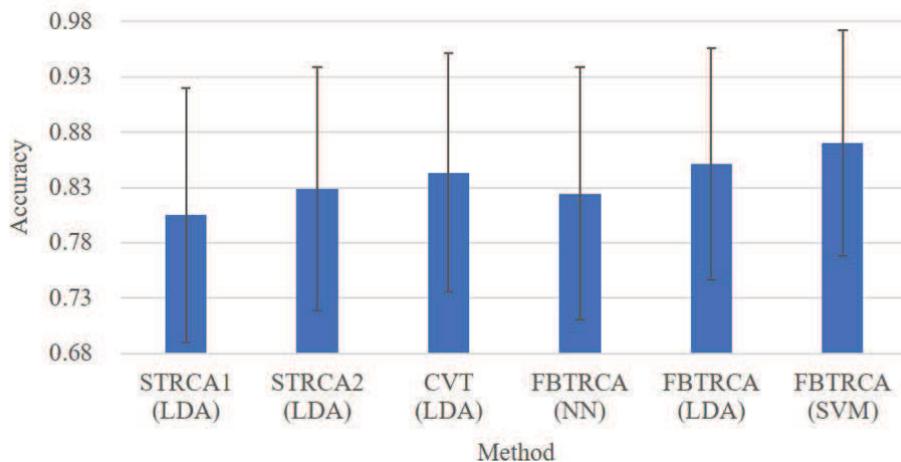}
    \caption{Accuracy Comparison against Benchmarks}
    \label{fig:3.7}
\end{figure}

In Figure \ref{fig:3.7} and Table \ref{tab:3.5}, the accuracies of these methods are listed. All of these methods have similar standard deviations. The proposed FBTRCA classified with SVM has the best mean value and standard deviation in comparison to the benchmarks and FBTRCA classified with LDA or NN classifiers.

\begin{table}[htbp]
    \centering
    \caption{Accuracy Comparison against Benchmarks}
    \footnotesize
    \begin{tabular}{c|c|c|c|c|c|c}
        \toprule
        \multirow{2}{*}{Method} & STRCA1 & STRCA2 & CVT & FBTRCA & FBTRCA & FBTRCA \\
        & (LDA) & (LDA) & (LDA) & (NN) & (LDA) & (SVM) \\
        \midrule
        mean value & 0.8048 & 0.8287 & 0.8431 & 0.8246 & 0.8514 & \textbf{0.8700} \\
        \midrule
        standard deviation & 0.1149 & 0.1101 & 0.1078 & 0.1141 & 0.1046 & \textbf{0.1022} \\
        \bottomrule
    \end{tabular}
    \label{tab:3.5}
\end{table}

\begin{figure}
    \centering
    \subfigure[Feature Distribution in Training Set]{\includegraphics[width=0.45\textwidth]{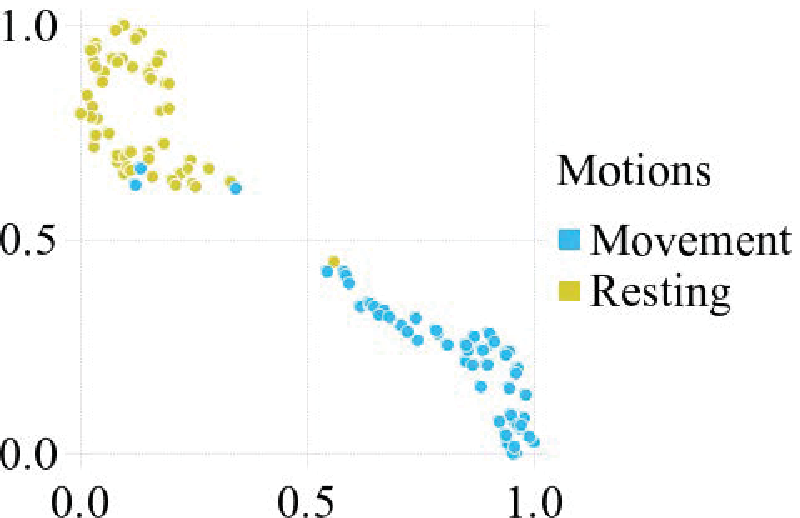}}
    \subfigure[Feature Distribution in Testing Set]{\includegraphics[width=0.45\textwidth]{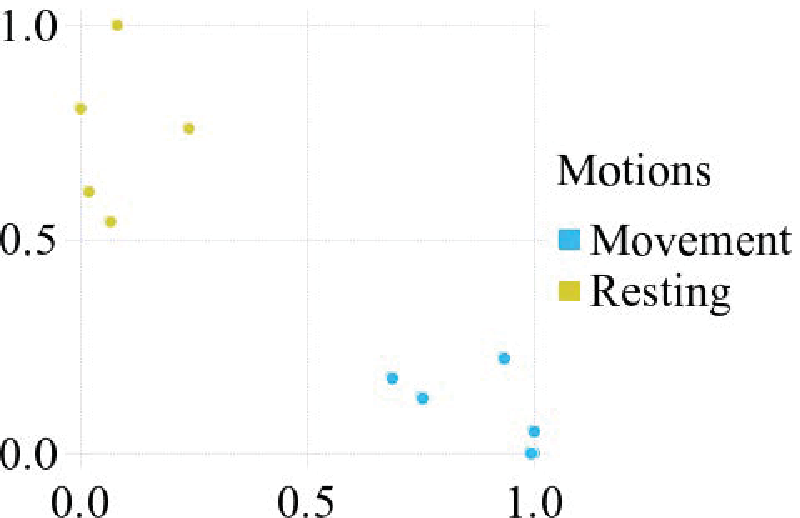}}
    \caption{Feature distributions of subject 1 in UMAP: (a) training set (b) testing set. In the UMAP settings, the neighbor number is set to 15 and the minimal distance is set to 0.001. The UMAP model is trained with 200 epochs. The features in the training set are first used to train the UMAP model and the embedding of the training set is obtained. Then the trained UMAP model is applied to the testing set to transform the embedding of the testing set. The embedding of UMAP is normalized with minimal-maximal normalization before visualization in this figure.}
    \label{fig:3.8}
\end{figure}

Uniform manifold approximation and projection (UMAP) is a tool that can visualize the high-dimensional features in a two-dimensional view \cite{2018UMAP}. In Figure \ref{fig:3.8}, the feature distributions extracted with the proposed FBTRCA method are given. The feature distributions consist of the distributions of both the training set and testing set of subject 1 in the classification between \textit{elbow flexion} and \textit{rest}. The feature points of the two classes are well separated with FBTRCA in both the training set and testing set.

\section{Discussion}
\label{sec:discuss}
In this study, we proposed a new pre-movement decoding method, FBTRCA. The FBTRCA method is proposed to solve the problems that STRCA has difficult decoding pre-movement patterns. The accuracies of STRCA are unstable in the subbands of the low-frequency domain. It is difficult to select a proper frequency range for each subject due to individual differences. Thus, the STRCA cannot achieve stable and satisfying performance. FBTRCA solves this problem with a filter bank technique, which is used in the FBCSP method (MI analysis) \cite{Ang2008} and the FBCCA method (SSVEP analysis) \cite{Chen_2015}. Both FBCSP and FBCCA are developed by incorporating an original EEG processing method such as CSP or CCA with the filter bank technique. During the development of FBTRCA, there are two assumptions:

\noindent (1) The STRCA is a simple and efficient method in pre-movement decoding, such as the CSP in MI analysis or CCA in SSVEP analysis;

\noindent (2) There are subbands of a certain frequency range setting in which STRCA achieves comparable and acceptable accuracies.

In the first assumption, STRCA is supposed to have a simple structure and show good accuracy performance in pre-movement decoding. In previous applications equipped with the filter bank technique, FBCSP or FBCCA was developed from the classical CSP or CCA method. The standard CSP method consists of a spatial filter and the logarithmic variance features. The standard CCP method is even simpler, which directly measures canonical correlation between the EEG signals and the reference signals. Although the structure of CSP or CCA is simple, both methods play an important role in the detection of motor imagination and visual stimuli. The STRCA also has a simple structure that consists of a spatial filter and six correlation coefficient features and has an acceptable performance even in comparison to SSSF \cite{duan_decoding_2021,jeong_decoding_2020}. Because of the simple structure and stable classification performance, STRCA features extracted from each subbands are expected to represent the main patterns in the corresponding frequency range.

In the second assumption, STRCA is supposed to decode pre-movement patterns in subbands of a frequency range setting. STRCA can decode pre-movement patterns in the low-frequency domain, such as the frequency range 0.05$\sim$10 $Hz$. However, the classification accuracies are unknown for STRCA in the subbands of 0.05$\sim$10 $Hz$, such as 8$\sim$9 $Hz$. In the filter bank technique, various subbands are required from which the extracted features can achieve an acceptable classification performance. In MRCP analysis, such a frequency range setting has not been determined. Therefore, three frequency range settings are first explored in Section \ref{ssec:frs}.

To incorporate the filter bank technique into the STRCA, three steps are adopted in the design of FBTRCA:

\noindent (1) Select the frequency range setting;

\noindent (2) Decide on how to select features; and

\noindent (3) Classify essential features with classifiers.

First, in the designation of filter banks, three frequency range settings are compared, including $M_1$, $M_2$ and $M_3$. According to the result analysis in each subband of three frequency settings, the accuracies of the subbands in $M_3$ have comparable and acceptable performance. The possible reason is that the low cut-off frequency of $M_3$ covers a extremely low frequencies. The bandpass filter used here filters the direct components. However, the EEG signals at low frequencies still need to be maintained. To obtain more features in the low-frequency domain, the low cut-off frequencies of the subbands in $M_3$ are shifted slightly from 0.5 $Hz$ to 0.05 $Hz$ with step 0.05 $Hz$.

Second, after extracting CCP features from subbands of $M_3$, the essential features are extracted from these features with feature selection methods. Here, we simultaneously consider the feature arrangement type and the feature selection method. The mutual-information based feature selection methods are compared, including MIQ, MRMR, MAXREL, MINRED, QPFS, CIFE, CMIM and MRMTR. The parameters K1 and K2 of two arrangement types are decided for eight selection methods.

Third, three binary classifiers are compared, including LDA, SVM and NN. In STRCA, LDA is used to classify STRCA features because the number of CCP features is six. LDA is efficient in dealing with such data with fewer features. In FBTRCA, the feature number is greater than TRCA. The SVM classifier shows a better performance in this study.

With the analysis of the three steps above, FBTRCA successfully incorporates the filter ban technique into the STRCA method. However, because FBTRCA is developed based on STRCA and aims to improve the classification performance of STRCA, FBTRCA inherits the application scenario from STRCA in which the FBTRCA is insufficient to detect the movement onset online.

\section{Conclusion}
\label{sec:conclu}
The proposed FBTRCA incorporates the filter bank technique and solves the unstable accuracy problem. There are four steps in FBTRCA. First, EEG signals are divided into various subbands in the low-frequency domain with frequency range setting $M_1$. Second, STRCA features are extracted from these subbands. Third, the mutual-information based feature selection method, MRMR, is used to select essential features from all of these CCP features. Finally, the essential features are classified with the binary classifier SVM. In comparison to the accuracy of STRCA (0.8287$\pm$0.1101) or the CVT method (0.8431$\pm$0.1078), the proposed FBTRCA method achieves a higher and more stable performance (0.8700$\pm$0.1022).

\newpage
\bibliographystyle{unsrt}
\bibliography{main}
\end{document}